\documentclass[aps,prl,twocolumn,superscriptaddress]{revtex4-1}
\usepackage{graphicx}
\usepackage{amssymb}
\usepackage{amsmath}
\usepackage{bm}
\usepackage{xcolor}

\begin{document}

\title{Observation of Two Sound Modes in a Binary Superfluid Gas}

\author{Joon Hyun Kim}
\affiliation{Department of Physics and Astronomy, and Institute of Applied Physics, Seoul National University, Seoul 08826, Korea}

\author{Deokhwa Hong}
\affiliation{Department of Physics and Astronomy, and Institute of Applied Physics, Seoul National University, Seoul 08826, Korea}

\author{Y. Shin}
\email{yishin@snu.ac.kr}
\affiliation{Department of Physics and Astronomy, and Institute of Applied Physics, Seoul National University, Seoul 08826, Korea}
\affiliation{Center for Correlated Electron Systems, Institute for Basic Science, Seoul 08826, Korea}

\date{\today}

\begin{abstract}
We study the propagation of sound waves in a binary superfluid gas with two symmetric components. The binary superfluid is constituted using a Bose-Einstein condensate of $^{23}$Na in an equal mixture of two hyperfine ground states. Sound waves are excited in the condensate by applying a local spin-dependent perturbation with a focused laser beam. We identify two distinct sound modes, referred to as density sound and spin sound, where the densities of the two spin components oscillate in phase and out of phase, respectively. The observed sound propagation is explained well by the two-fluid hydrodynamics of the binary superfluid. The ratio of the two sound velocities is precisely measured with no need for absolute density calibration, and we find it in quantitatively good agreement with known interaction properties of the binary system.   
\end{abstract}

\maketitle

Sound propagation in a superfluid is a characteristic transport phenomenon revealing the microscopic and thermodynamic properties of the superfluid system. In a long-wavelength limit, a sound wave propagates without distortion, reflecting the linear dispersion of the gapless Goldstone excitation mode of the superfluid~\cite{Landau41}. At finite temperatures, containing a normal fluid within it, a superfluid system supports two types of sound waves, referred to as first and second sounds~\cite{Fairbank47,Stringari13}, and their propagation speeds are determined as functions of thermodynamic quantities such as superfluid density, entropy density, and compressibility~\cite{Lee59}. Thus, the study of sound propagation inclusively measures our understanding of the superfluid system.

A superfluid with two superflowing components has been studied with great interest since the discovery of $^3$He--$^4$He mixtures~\cite{Yaqub65}, and this fascination is further motivated by recent experimental developments in ultracold atomic gas mixtures~\cite{Modugno02,Salomon15,Tarruell18}, exciton-polariton condensate~\cite{Lagoudakis09}, and two-gap superconductors~\cite{Zehetmayer13}. In the binary superfluid system, interactions between two components induce the mixing of the low-energy excitation modes of each component, giving rise to two new hybridized sound waves. A peculiar property of the binary system is that the superflow of one component can experience a nondissipative drag from the movement of the other component~\cite{Bashkin75}, known as the Andreev-Bashkin (AB) effect. Such a mutual entrainment influence affects the speed of sound and can be important in the stability and robustness of the superfluidity of the binary system, particularly in a strongly interacting regime~\cite{Kobyakov17}. The quantitative understanding of the AB effect in a binary superfluid has been attentively pursued in recent theoretical works~\cite{Shevchenko05,Recati17,Andreev18,Giorgini18,Enss19}.

\begin{figure}
	\includegraphics[width=8.5cm]{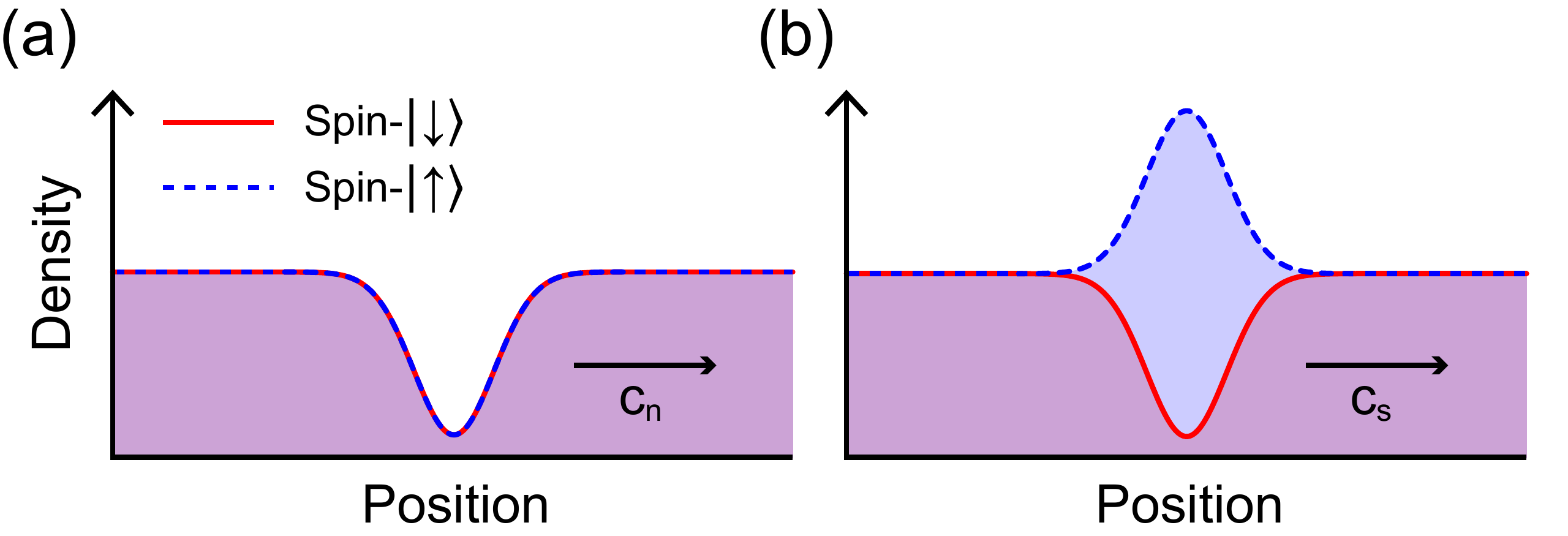}
	\caption{Sound waves in a symmetric binary superfluid. (a) Ordinary density sound where the two superfluid components oscillate in phase and (b) spin sound wave where they oscillate out of phase. The red solid and blue dashed lines indicate the spatial density profiles of the two components. $c_n$ and $c_s$ denote the propagation speeds of the density and spin sounds, respectively.}
\end{figure}

In this paper, we investigate sound propagation in a binary superfluid gas as constituted by a Bose-Einstein condensate (BEC) of $^{23}$Na in an equal mixture of two hyperfine ground states. Here the two components are identical in mass, density, and intra-component interactions. From the $\mathbb{Z}_2$ symmetry, it is expected that the superfluid shows two modes of density oscillations, i.e., sound waves, where the densities of the two components oscillate in phase and out of phase, respectively (Fig.~1). The in-phase oscillations correspond to an ordinary pressure wave, whose propagation speed is determined by the compressibility of the system. Meanwhile, the out-of-phase oscillations are a wave of the density difference between the two components, referred to as $\textit{spin}$ sound, regarding the two components as two opposite spin states, $|$$\uparrow\rangle$ and $|$$\downarrow\rangle$~\cite{Pavloff14,Zhang15,Kamchatnov18}. We observe two distinct sound waves propagating with different speeds in the condensate and identify the fast wave with ordinary density sound and the slow wave with spin sound. The measured propagation speeds of the two sound waves are found to be well explained by the two-fluid hydrodynamics of the binary superfluid and the ratio of the two sound velocities is precisely determined with no need of absolute density calibration. This work opens a quantitative study of the sound propagation in the binary superfluid system.

Our experiment starts with the preparation of a BEC of $^{23}$Na in the $|F$$=$$1,m_{F}$$=$$0\rangle$ hyperfine state~\cite{Choi12}. The condensate, typically containing $\approx 3\times10^{6}$ atoms, is trapped in an optical dipole trap with trapping frequencies of $(\omega_{x},\omega_{y},\omega_{z}) = 2\pi\times(5.4,8.0,571)$~Hz and its Thomas-Fermi radii are $(R_{x},R_{y},R_{z})\approx(168,113,1.6)~\mu$m. A binary superfluid is realized by transferring the atoms in the $|m_F$$=$$0\rangle$ state to a superposition of the two spin states, $|$$\uparrow\rangle \equiv |m_F$$=$$1\rangle$ and $|$$\downarrow\rangle \equiv |m_F$$=$$-1\rangle$, by applying a short radio-frequency (rf) pulse so that the condensate comprises an equal mixture of the two spin components. The $s$-wave scattering lengths for the intra- and inter-component collisions are given by $a=54.54(20)a_0$ and $a_{\uparrow\downarrow}=50.78(40)a_0$, respectively, with $a_0$ being the Bohr radius~\cite{Oberthaler11}. The two components are miscible for $a>a_{\uparrow\downarrow}$~\cite{Ketterle98}. To prevent the spin-exchange process generating the $|m_F$$=$$0\rangle$ component, we tune the quadratic Zeeman energy to $q/h\approx-5.0$~Hz using microwave field dressing~\cite{Bloch06,Liu14}, and the binary superfluid is stablized. The thermal fraction of the sample is less than $10~\%$. The external magnetic field is $B_z=50$~mG and its gradient is controlled to be less than 0.1~mG/cm~\cite{Kim19}.

Sound waves are generated by applying local potential perturbations on the condensate using a focused Gaussian laser beam, as in previous works~\cite{Ketterle97,Thomas07,Straten09,Beugnon18}. The key feature of our experiment is that the magnitude of the optical potentials $V^0_{\uparrow(\downarrow)}$ for the spin--$\uparrow$($\downarrow$) components can be differentiated by using a near-resonant laser beam to generate spin-dependent perturbation~\cite{Lebrat19}. The laser beam penetrates along the $z$ axis at the center of the condensate. We adiabatically turn on the laser beam by increasing its intensity for 200~ms and then rapidly switch it off in 1~ms. The generation of sound waves and their subsequent propagation are probed by taking absorption or spin-sensitive phase-contrast imaging at various hold times $t$ after turning off the laser beam.

\begin{figure}[t]
	\includegraphics[width=8.5cm]{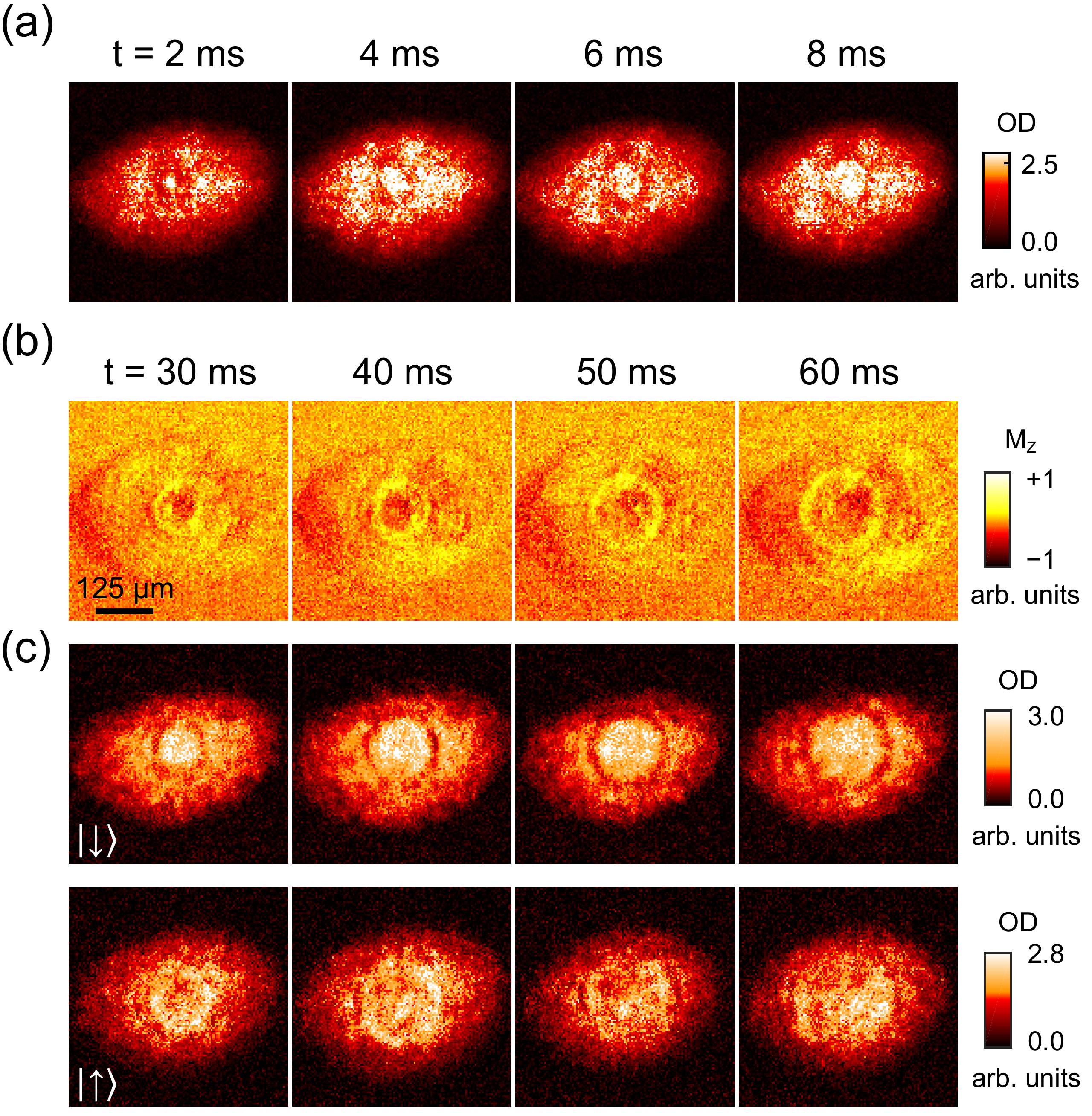}
	\caption{Observation of the sound propagation in a binary superfluid. (a) $in~situ$ optical density (OD) images of the two-component Bose-Einstein condensate at various hold times $t$ after applying a local density perturbation at the center of the condensate. A ring-shaped rarefaction pulse is generated, and it propagates outwardly. (b) $in~situ$ magnetization ($M_z$) images of the condensate at various $t$ after applying a local spin perturbation and (c) the corresponding OD images of the two spin components taken after a 19-ms time of flight with Stern-Gerlach (SG) spin separation. The spin-$\downarrow$ component shows a density dip and the spin-$\uparrow$ component a density lump at wave pulse position. 
	}
\end{figure}

We first investigate a density perturbation case with $V^0_\uparrow=V^0_\downarrow$, where we use a 532-nm far-detuned laser beam with identically repulsive dipole potentials for the two spin states. The $1/e^2$-intensity radius of the laser beam is 18.3(3)$~\mu$m, and its potential height before turning it off is approximately five times the chemical potential $\mu$ of the sample, resulting in a density-depleted hole in the condensate. In Fig.~2(a), the {\it in situ} density images of the sample for various hold times are displayed. We observe that a ring-shaped rarefaction pulse is generated, and it propagates out in a radial direction. Taking absorption imaging after Stern-Gerlach (SG) spin separation~\cite{Choi12}, we find that both spin components show density dips at the pulse position, which indicates that the generated wave pulse is a sound wave in the in-phase oscillating mode. During its propagation, the spatial width of the wave pulse is maintained at $\approx 23~\mu$m, implying a dispersionless characteristic.

Next, we investigate the sound wave generation by spin-dependent perturbation with $V^0_\uparrow=-V^0_\downarrow$, where we employ a 589-nm near-resonant laser beam. The optical frequency of the laser beam is tuned at $2\pi\times508.505$~THz between the $D1$ and $D2$ lines of $^{23}$Na to generate asymmetric optical potentials for the two spin states. The laser beam is set to be $\sigma^-$--polarized and the resulting optical potential is attractive (repulsive) for the spin--$\uparrow$ ($\downarrow$) component. The $1/e^2$ radius of the laser beam is 11(1)$~\mu$m and its potential magnitude is $V^0_{\uparrow(\downarrow)} \approx \mp 1.1\mu$.

Figure~2(b) shows the $in~situ$ magnetization images of the sample for various hold times after turning off the 589-nm laser beam~\cite{Seo15}. A ring-shaped wave pulse carrying a positive magnetization is observed as propagating out from the center of the condensate. It is noticeable that the magnetization pulse propagates much more slowly than the density wave pulse presented in the previous case, indicating that it is a different type of sound wave. From the images taken after SG spin seperation [Fig.~2(c)], we observe that the spin--$\downarrow$ component shows a density dip at the position of the propagating wave pulse whereas the spin--$\uparrow$ component shows a density lump at the same position. This clearly demonstrates the generation of a spin sound wave in the binary superfluid system.

\begin{figure}[t]
	\includegraphics[width=8.5cm]{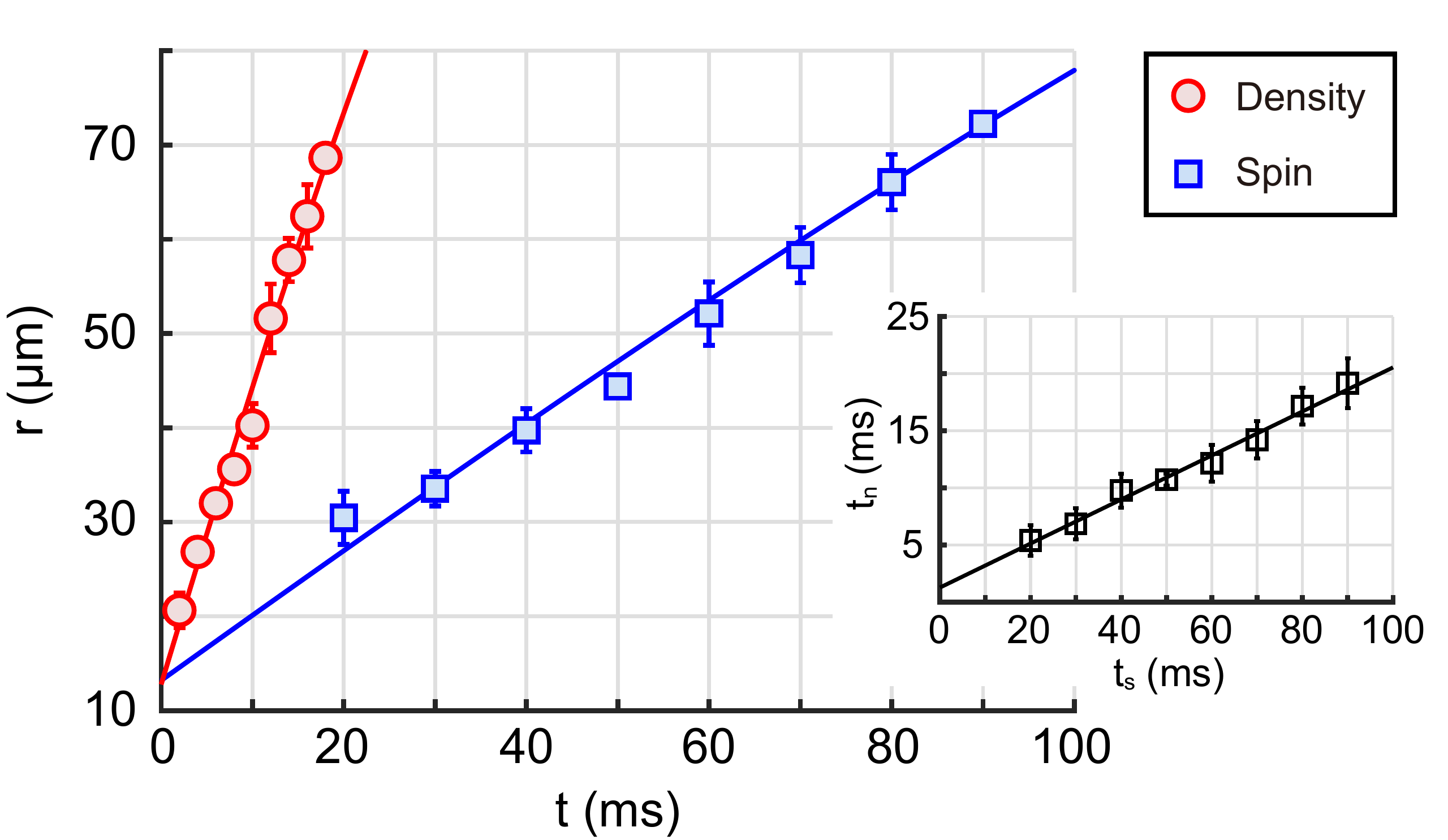}
	\caption{Time evolution of the radius of the ring-shaped wave pulse; $r_n$ (red circles) for the density sound wave and $r_s$ (blue squares) for the spin sound wave. Each data point consists of five measurements of the same experiment and its error bar indicates their standard deviation. The solid lines are the model function fit to the experimental data. The inset shows the constructed data $\{t_s, t_n\}$ for the scale analysis of the sound wave propagation (see the text for details) and the solid line is a linear function fit to the data.
	}
\end{figure}

In a long-wavelength limit, zero-temperature dynamics of a binary superfluid is well described by two-fluid hydrodynamic equations~\cite{Kamchatnov18}: 
\begin{eqnarray}
\partial_t n_i &+&\nabla \cdot (n_i \bm{u}_i) =0 \nonumber \\
\partial_t \bm{u}_i &+& \nabla (\frac{1}{2} u_i^2 + \frac{g}{m} n_i + \frac{g_{\uparrow\downarrow}}{m} n_j) =0,
\end{eqnarray}
where $\{n_i,\bm{u}_i\}$ is the density and velocity of the spin-$i$ component ($i,j=\uparrow, \downarrow$ and $i\neq j$), $g_{(\uparrow\downarrow)}=\frac{4\pi \hbar^2}{m} a_{(\uparrow\downarrow)}$, and $m$ is the atomic mass. The first equation is the continuity equation and the second is the Euler equation with the chemical potential of spin-$i$ component, $\mu_i=g n_i + g_{\uparrow \downarrow} n_j$. By linearizing the equations for a stationary state with $\bm{u}_{\uparrow(\downarrow)}=0$ and finding the condition to have a traveling wave solution of $\delta n_i=A_i \sin(\bm{k}\cdot \bm{r}-\omega t)$, we obtain the wave velocity $v_s=\omega/k$ as
\begin{eqnarray}
v_{s\pm}^2&=&\frac{1}{2m}\Big[ g n \pm \sqrt{g_{\uparrow\downarrow}^2 n^2  + (g^2 - g_{\uparrow\downarrow}^2)n_s^2 }~\Big],
\end{eqnarray}
where $n=n_\uparrow+n_\downarrow$ is the total density and $n_s=n_\uparrow-n_\downarrow$ is the spin density. For $g>g_{\uparrow\downarrow}>0$, there are always two sound modes with different propagation speeds. 

In a symmetric binary superfluid with $n_s=0$, the two sound velocities are given by $v_{s\pm}=\sqrt{\frac{1}{2m}(g \pm g_{\uparrow\downarrow})n}$. The fast wave is density sound for $A_\uparrow=A_\downarrow$ and the slow wave is spin sound for $A_\uparrow=-A_\downarrow$, which are consistent with our experimental observations. Mass current $\bm{j}_m=n_\uparrow \bm{u}_\uparrow + n_\downarrow \bm{u}_\downarrow$ and spin current $\bm{j}_s=n_\uparrow \bm{u}_\uparrow - n_\downarrow \bm{u}_\downarrow$ are decoupled in the superfluid, which reflects a peculiar consequence of the $\mathbb{Z}_2$ symmetry of the binary system, and the Bogoliubov quasiparticles are thus given by pure phonons and magnons~\cite{Ueda12}. Our observation of the spin sound propagation demonstrates the existence of the gapless magnon mode associated with the relative phase of the two spin components and corroborates the spin superfluidity of the binary system~\cite{Malpuech13,Kim17,Ferrari18}.

We measure the propagation speed of the sound wave from the time evolution of the radius $r$ of the ring-shaped wave pulse (Fig.~3). Here, $r$ is determined from the {\it in situ} absorption and magnetization images for the density and spin sounds, respectively [Fig.~2(a) and 2(b)]. Taking into account the inhomogeneous density profile of the trapped sample, the radial dependence of the propagation speed is modelled as $v_r(r)=c\sqrt{1-r^2/R^2}$ with the peak speed $c$ at the center and $R=\sqrt{R_xR_y}$. By integrating $v_r(r)$ over time, we obtain a model function of the ring radius as $r(t)=R \sin (ct/R +\theta_0)$ and we determine the speed $c$ and the initial position $r(0)=R \sin(\theta_0)$ of the sound wave from the model function fit to the experimental data. In our measurements, the sound velocities are given by $c_n=3.23(18)$~mm/s for density sound and $c_s=0.70(4)$~mm/s for spin sound, which are found to be in agreement with the estimations of $v_{s+}=3.22(5)$~mm/s and $v_{s-}=0.61(4)$~mm/s from Eq.~(2). For the calculation of $v_{s\pm}$, we use the effective density $n=(2/3)n_0$ with $n_0$ being the peak density of the condensate, under the assumption of the hydrodynamics equilibrium along the tight confining $z$ direction of the highly oblate sample~\cite{Zaremba98,Stringari98,Moritz15}. The initial positions $r_{n(s)}(0)$ of the sound waves are measured to be $\approx 13~\mu$m, which are comparable to the spatial sizes of the used laser beams.

Here we remark that, although the individual velocities of $c_n$ and $c_s$ are dependent on the details of the sample condition such as the particle density and the trapping geometry, the ratio $\gamma=c_s/c_n$ is faithfully determined only by the intrinsic properties of the binary superfluid because both velocities are measured in the same sample condition. Indeed, $\gamma$ can be directly measured from a scale analysis of the experimental data for the two sound waves without any assumption concerning the density profile of the sample. We perform the scale analysis as follows; for each data point $\{t_s, r_s\}$ in the spin sound measurement we determine the corresponding time $t_n$ for the density sound wave to have the same radius $r_n=r_s$ from linear interpolation of the density sound measurement data. We then measure $\gamma$ by fitting a linear function $t_n=\gamma t_s + \beta$ to the constructed data $\{t_s, t_n\}$ (Fig.~3 inset). Here the offset $\beta$ accounts for the difference of the initial positions of the two sound waves. Our measurement gives $\gamma=0.193(22)$, which is in quantitatively good agreement with the estimated value of $\gamma_0=\frac{v_{s-}}{v_{s+}}=\sqrt{\frac{a-a_{\uparrow\downarrow}}{a+a_{\uparrow\downarrow}}}=0.189(11)$.

Because $\gamma$ can be measured without absolute density calibration, we might consider a precise measurement of $\gamma$ for probing subtle interaction effects, such as the AB entrainment effect. In Ref.~\cite{Recati17}, it was shown that the AB effect modifies only the speed of spin sound, not density sound, meaning that the superfluid drag effect appears as a shift in $\gamma$. In our weakly interacting system with $n_0 a^3 \approx 1.4\times 10^{-6}$, the fractional weight of the superfluid drag is predicted to be very small, $\approx 4\times 10^{-4}$~\cite{Shevchenko05}, and its detection is limited by the current uncertainties of the scattering lengths, $a$ and $a_{\uparrow\downarrow}$. For the quantitative interpretation of $\gamma$, there are additional factors to be taken into account. One is the effect of the Lee-Huang-Yang (LHY) corrections arising from quantum fluctuations. For the given gas parameter, the relative shift of $c_n$ due to the LHY corrections is estimated to be about 0.5\%~\cite{Ueda10}, which is one order of magnitude larger than the AB effect in $c_s$. Another factor is finite temperature effects. We note, in particular, that in our highly oblate sample, the density and spin healing lengths for the peak density $n_0$ are given by $\xi_{n}\approx 0.5~\mu$m and $\xi_{s}\approx 2.6~\mu$m, respectively, so $\xi_{n}<R_z<\xi_{s}$. Thus, thermal effects might have different dimensional characteristics for the density and spin channels, possibly resulting in a variation of $\gamma$.

\begin{figure}[t]
	\includegraphics[width=8.5cm]{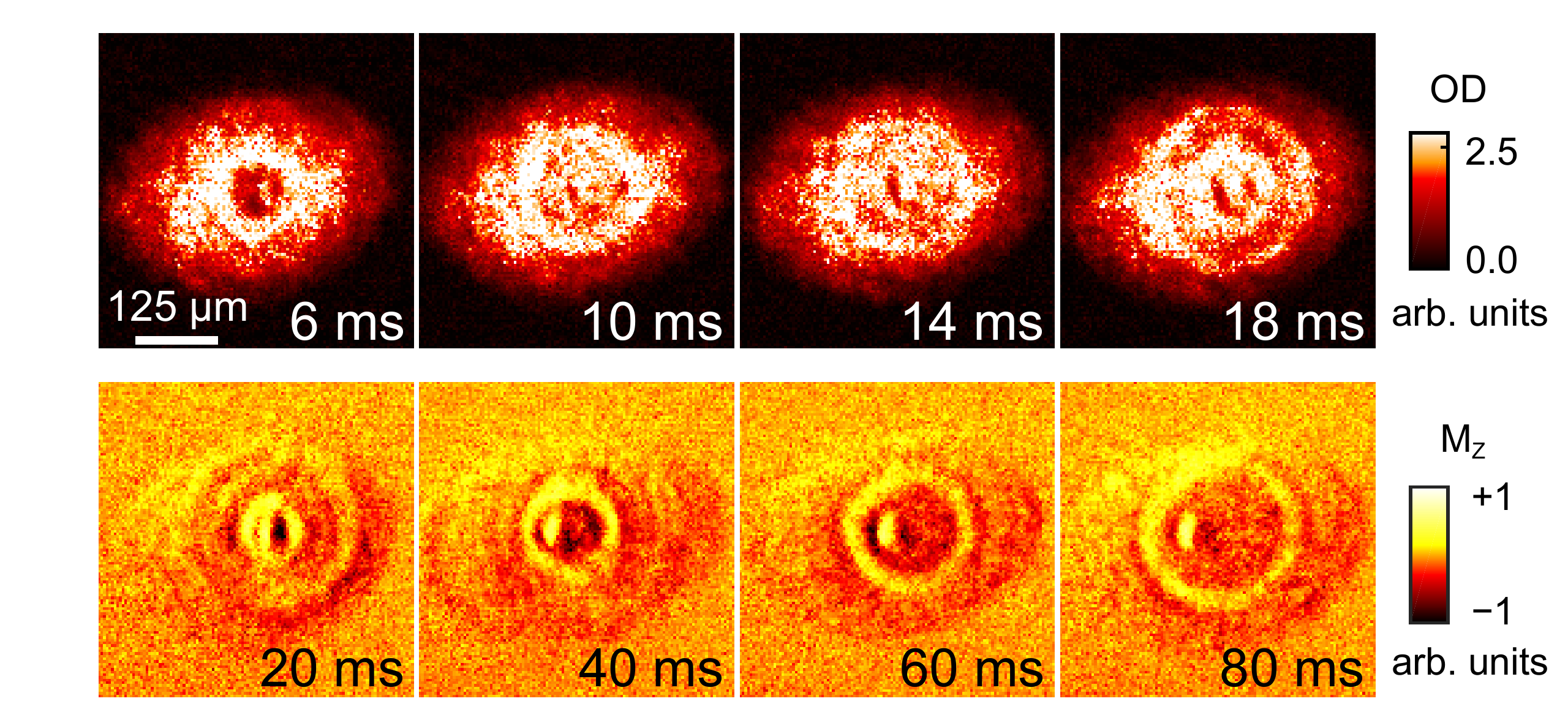}
	\caption{Simultaneous generation of density and spin sound waves with strong spin-dependent perturbation. OD images (upper row) and magnetization images (lower row) for various hold times, taken with a 19-ms time of flight. A fast density wave and a slow spin wave propagate radially from the center. A solitonic object with a magnetized core is created together with the two sound waves, moving slower than the spin wave.
	}
\end{figure}

Finally, we explore the case of $|V^0_\uparrow|\neq|V^0_\downarrow|$ for generating both density and spin sound waves simultaneously. We survey various perturbation regimes by changing the frequency and intensity of the near-resonant laser beam. As a result, when $V_\uparrow\approx-3\mu$ and $V_\downarrow\approx \mu$, we observe that both density and spin sound waves are generated~(Fig.~4). They propagate separately in the condensate, highlighting their different speeds. The density wave appears as a density lump because $V_\uparrow+V_\downarrow <0$ and the spin sound wave carries positive magnetization because $V_\uparrow-V_\downarrow<0$. 

A notable observation is that a soliton-like magnetized object is created together with the two sound waves. The magnetized object moves slower than the spin wave, with preserving its spatial size and shape (Fig.~4). We interpret it as a small dipole of half-quantum vortices (HQVs) with same core magnetization, which explains the bending shape of the object and its linear motion~\cite{Seo16}. HQVs are the defects involving both mass and spin circulations, so their creation indicates that there was nonlinear coupling between mass and spin currents in the sound wave generation with the strong spin-dependent perturbation. The dipole's moving velocity of $\approx c_s/2$ suggests that the separation between the two HQVs is $\approx 2\xi_s$, implying that their magnetized cores are almost coalesced~\cite{Seo15}. In the experiment, the generation of the solitonic object was quite deterministic and we believe that it must be due to the asymmetric shape of the optical potential.

In conclusion, we have investigated sound propagation in the symmetric binary superfluid and observed two distinct sound modes in density and spin channels, respectively. The ratio $\gamma$ of the two sound velocities was precisely measured with no need for absolute density calibration. An interesting extension of this work would be to investigate the temperature dependence of $\gamma$ and its evolution in the dimensional crossover to 2D~\cite{Andreev18} and 1D~\cite{Giorgini18}. We expect that the spin-dependent potential of the near-resonant laser beam can be extensively used to investigate dynamics of various topological objects in the spinor superfluid, such as HQVs~\cite{Seo16}, skyrmions/merons~\cite{Choi12}, and magnetic solitons~\cite{Stringari16}.

\begin{acknowledgments}
This work was supported by the Samsung Science and Technology Foundation under Project Number SSTF-BA1601-06.
\end{acknowledgments}

\end{document}